\documentclass[twoside,12pt]{article}
\usepackage{amsmath}
\usepackage{amssymb}
\usepackage{latexsym}
\usepackage{epsfig}
\usepackage{cite}
\newlength{\dinwidth}
\newlength{\dinmargin}
\newtheorem{theorem}{Theorem}[section]
\newtheorem{prop}[theorem]{Proposition}
\newtheorem{lemma}[theorem]{Lemma}
\newtheorem{cor}[theorem]{Corollary}
\newtheorem{definition}{Definition}
 
\newenvironment{proof}{\medskip \noindent 
            {\bf Proof.}}{ \hfill $\square$ \medskip}

\newcommand{\wm}{{W\!\!\!\!\text{\raisebox{1.8ex}[0ex][0ex]%
{$\scriptscriptstyle\sim$}}}}
\def\wMin{{\Ws\!\!\!\!\text{\raisebox{1.7ex}[0ex][0ex]%
{$\scriptscriptstyle\sim$\,}}}}

\def\WRM{{\wm}\,^{(1)}} 
\def\EM{{E\!\!\!\text{\raisebox{1.8ex}[0ex][0ex]%
{$\scriptscriptstyle\sim$\,}}}}
\def\ERM{\EM^{(1)}} 
\newcommand{\Wtilde}{{\wm}} 
\newcommand{\jm}{{j\!\!\text{\raisebox{1.6ex}[0ex][0ex]%
{$\scriptscriptstyle\sim$\!}}}}
\newcommand{\lambdam}{{\lambda\!\!\!\text{\raisebox{1.7ex}[0ex][0ex]%
{$\scriptscriptstyle\sim$}}}}
\newcommand{\gammam}{{\gamma\!\!\!\text{\raisebox{1.3ex}[0ex][0ex]%
{$\scriptscriptstyle\sim$}}}}
\newcommand{\rhom}{{\rho\!\!\!\text{\raisebox{1.25ex}[0ex][0ex]%
{$\scriptscriptstyle\sim$}}}}
\def\LM{{L\!\!\!\text{\raisebox{1.9ex}[0ex][0ex]%
{$\scriptscriptstyle\sim$\,}}}}

\newcommand{\wrw}{{W\!\!\!\!\!\text{\raisebox{-0.7ex}[0ex][0ex]%
{$\scriptscriptstyle\sim$\,}}}}
\newcommand{\wrwsub}[1]{\wrw{}_{#1}}
\def\dcrw{{\Os\!\!\!\!\text{\raisebox{-0.7ex}[0ex][0ex]%
{$\scriptscriptstyle\sim$\,}}}} 
\newcommand{\dcrwsub}[1]{\dcrw{}_{#1}}   
\def\wRW{{\Ws\!\!\!\!\!\text{\raisebox{-0.7ex}[0ex][0ex]%
{$\scriptscriptstyle\sim$ }}}} 
\def\Arw{{\As\!\!\!\!\text{\raisebox{-0.8ex}[0ex][0ex]%
{$\scriptscriptstyle\sim$\,}}}} 
\def\Rrw{{\Rs\!\!\!\!\text{\raisebox{-0.7ex}[0ex][0ex]%
{$\scriptscriptstyle\sim$\,}}}}  
\def\wnetrw{\{\Arw(\wrw)\}_{\wrw \in \wRW}}
\def\wrnetrw{{\{\Rrw(\wrw)\}_{\wrw \in \wRW}}}
\def\orw{{\omega\!\!\!\!\text{\raisebox{-0.7ex}[0ex][0ex]%
{$\scriptscriptstyle\sim$\,}}}}
\def\Orw{{\Omega\!\!\!\!\text{\raisebox{-0.7ex}[0ex][0ex]%
{$\scriptscriptstyle\sim$\,}}}} 
 
\def\grw{{g\!\!\!\text{\raisebox{-1.05ex}[0ex][0ex]%
{$\scriptscriptstyle\sim$\,}}}} 
\def\gammarw{{\gamma\!\!\!\text{\raisebox{-1.0ex}[0ex][0ex]%
{$\scriptscriptstyle\sim$\,}}}} 
\def\nrw{{\nabla\!\!\!\!\text{\raisebox{-0.7ex}[0ex][0ex]%
{$\scriptscriptstyle\sim$\,}}}{}} 

\def\erw{{E\!\!\!\!\text{\raisebox{-0.7ex}[0ex][0ex]%
{$\scriptscriptstyle\sim$\,}}}} 
\def\frw{{F\!\!\!\!\text{\raisebox{-0.75ex}[0ex][0ex]%
{$\scriptscriptstyle\sim$\,}}}}

\def\wdS{{\Ws}}
\def\wnetds{{\{\As(W)\}_{W \in\wdS}}}
\def\wrnetds{{\{\Rs(W)\}_{W \in\wdS}}}
\def\erds{E^{(1)}}
\def\wrds{W^{(1)}}

\def\wedges{\Ws_{M}}
\def\wnet{{\{\As(W)\}_{W \in\wedges}}}
\def\wrnet{{\{\Rs(W)\}_{W \in\wedges}}}

\def\msc{{Modular Stability Condition}}

\def\As{{\mathcal A}}

\def\Hs{{\mathcal H}}

\def\Js{{\mathcal J}}

\def\Os{{\mathcal O}}

\def\Rs{{\mathcal R}}

\def\Ws{{\mathcal W}}

\def\Lid{SO_0(4,1)}

\def\idty{{\leavevmode\hbox{\rm 1\kern -.3em I}}}

\def\nind{\noindent}
\def\RR{\mathbb{R}}
\def\CC{\mathbb{C}}

\newcommand{\Min}{{$\RR^5$}}
\newcommand{\Minm}{{\RR^5}}
\newcommand{\RW}{\mbox{\it RW}\ }
\newcommand{\dS}{\mbox{\it dS}\ }
\newcommand{\RWm}{\mbox{\it RW}}
\newcommand{\dSm}{\mbox{\it dS}}
\newcommand{\nor}{{\rm nor}}
\newcommand{\pihalf}{{\frac{\pi}{2}}}
\newcommand{\half}{{\frac{1}{2}}}

\newcommand{\grad}{{\rm grad}}
\newcommand{\eps}{{\varepsilon}}
\newcommand{\epsc}{\hat{\eps}}

\newcommand{\artanh}{{\rm arctanh}}
\begin{document}
\title{Transplantation of Local Nets \\ and Geometric Modular Action\\
 on Robertson--Walker Space--Times}
\author{{Detlev Buchholz, Jens Mund}\\
Institut f\"ur Theoretische Physik, Universit\"at G\"ottingen,\\
Bunsenstra\ss e 9, D-37073 G\"ottingen, Germany\\
\vphantom{X}\\
{Stephen J.\ Summers } \\
Department of Mathematics, University of Florida,\\
Gainesville FL 32611, USA\\}

\date{\small Dedicated to Sergio Doplicher and John E.~Roberts} 

\maketitle 
{\abstract \nind A novel method of transplanting algebras of observables
from de Sitter space to a large class of Robertson--Walker
space--times is exhibited. It allows one to establish the existence
of an abundance of local nets on these spaces which comply with a
recently proposed condition of geometric modular action. The
corresponding modular symmetry groups appearing in these examples 
also satisfy a  condition of modular stability, which has been suggested 
as a substitute for the requirement of positivity of the energy in
Minkowski space. 
Moreover, they exemplify the conjecture that the modular symmetry
groups are generically larger than the isometry and conformal groups 
of the underlying space--times. 
}
\section{Introduction}

     An important problem in the theory of quantum fields on general
curved space--times is how to choose fundamental reference states which
could play a role similar to that of the vacuum state for quantum
fields on Minkowski space.\footnote{This problem was apparently first made
explicit in \cite{Full}.} Interesting suggestions have been made in
the recent past for selecting folia of physically relevant states
\cite{HNS,FH,Rad1,Rad2,BFK,Wald}. These approaches have in
common that they do not address the question of how to select a
fundamental reference state out of these folia.

In \cite{BS} a Condition of Geometric Modular
Action (henceforth, CGMA) was introduced in order to give a purely algebraic
selection criterion for such physically significant states on
arbitrary space--times. For a precise statement and brief discussion
of the CGMA, see Section~3. It was shown in~\cite{BDFS,F2}
in the special cases of four--dimensional Minkowski space and three--
and four--dimensional de Sitter space that from a state and net of
algebras satisfying the CGMA it is possible to derive the
isometry group of the respective space--time, along with a strongly
continuous unitary representation of this spacetime symmetry group
(for further developments, see \cite{BFS1,BFS2}). Moreover, the initial
state is invariant and the initial net of
algebras containing the observables of the theory is covariant under 
the action of this representation.  Hence, the spacetime symmetry
group and its action upon the observables of the theory were
derived from the observables and state and not
posited, as is customarily done.
Thus, the CGMA is indeed a distinctive feature of the states of
interest in these spaces~\cite{BW,TW}.

The primary purpose of this article is to suggest the CGMA's wider
range of applicability by providing examples of nets of algebras on a
class of Robertson--Walker space--times supplied with states which
together satisfy the CGMA. Equal\-ly significant, as we shall explain in
Section 3, is the fact that in these examples the modular symmetry
group is strictly larger than the isometry  groups of the
space--times. More specifically, the geometric action of the modular symmetry
group does  not in general implement point transformations.

The essential step in our analysis is of a purely geometric nature.
We shall exhibit maps $\Xi$ from a certain specific family of
regions in de Sitter space, called wedges, to a corresponding family
in the respective Robertson--Walker space--time. These maps are not
induced by point transformations, but nevertheless commute with the
operation of taking causal complements; moreover they induce an action
of the de Sitter group on their images.
This fact will enable us to transplant in a local and
covariant manner nets of algebras of observables affiliated with wedges in de
Sitter space to corresponding nets in the Robertson--Walker
spaces. If the underlying de Sitter theory also complies with the CGMA it
follows that the resulting Robertson--Walker theory has the
same property. This method of transplanting a local net from one
space--time to another is akin to the method of ``algebraic
holography'', by which Rehren has proved the Anti-de
Sitter -- conformal QFT correspondence~\cite{Re99}.

Our paper is organized as follows.
In Section 2 we outline the geometric background and exhibit the
pertinent properties of the maps $\Xi.$ Using these results, we
construct in Section 3  nets of algebras and corresponding states
on the specified class of Robertson--Walker space--times and establish
their desired modular properties.
Finally, in Section~4 we discuss the significance of our results.

\section{Geometric Considerations}
In this section we exhibit a natural correspondence between certain
specific families of causally closed regions (wedges) in a large class
of Robertson--Walker space--times. Moreover, we establish some
basic properties of this correspondence which enter in our
construction in the subsequent section.

Robertson--Walker space--times are Lorentzian warped products of a
connected open subset of $\RR$ with a  Riemannian manifold of
constant sectional curvature~\cite{BEE,HE,O}.
We restrict our attention here to the case of
positive curvature, which may be assumed to be $+1.$
The corresponding  Robertson--Walker space--times are homeomorphic
to $\RR \times S^3,$ and one can choose coordinates so that the metric
has the form
\begin{equation}
ds^2 = dt^2 - S^2(t) \, d\sigma^2.
\end{equation}
Here, $t$ denotes time, $S(t)$ is a strictly positive smooth function
and  $d\sigma^2$ is the time-independent metric on the unit sphere $S^3$:
\begin{equation}
d\sigma^2 = d\chi^2 + \sin^2 (\chi) \;
(d\theta^2 + \sin^2 (\theta) \, d\phi^2) .
\end{equation}
The isometry group for such space--times contains a subgroup isomorphic to
the rotation group $O(4)$. Indeed, for generic Robertson--Walker spaces, the
full isometry group is isomorphic to $O(4)$. 

Following \cite{HE} we define a rescaled time parameter $\tau(t)$ by
\begin{equation}
d\tau / dt = 1/S(t).
\end{equation}
     In terms of this new variable, the metric takes the form
\begin{equation} \label{1}
ds^2 = S^2(\tau)\big(d\tau^2 - d\sigma^2\big),
\end{equation}
where $S(\tau)$ is shorthand for $S(t(\tau))$.
Since $S$ is everywhere positive, $\tau$ is a continuous, strictly increasing
function of $t$; its range is therefore an open interval in $\RR$.
In this paper we restrict our attention to those Robertson--Walker space--times
with functions $S(t)$ such that the range of values of $\tau$ is
of the form $(-\alpha\frac{\pi}{2},\alpha\frac{\pi}{2})$,
with $\alpha \leq 1.$
We henceforth denote by $RW$ any one of this class of
Robertson--Walker space--times.

We mention as an aside that, adopting the
standard big bang model of a perfect fluid yields the Robertson--Walker
space--times as solutions of Einstein's field equations, which determine the
range of values of $\tau.$ In particular, if the
pressure and cosmological constant equal 0, one has the indicated
range of values for $\tau$ with $\alpha=1$;
if the pressure is strictly positive, one has the case
$\alpha<1$~\cite{HE}.

     As is well known, the four--dimensional de Sitter space--time $dS$
can be embedded into five--dimensional Minkowski space $\RR^5$ as follows:
\begin{equation}  \label{eqdSMin}
dS = \{ x \in \RR^5 \mid x_0^2 - x_1^2 - \ldots - x_4^2 = -1 \} .
\end{equation}
This space--time is topologically equivalent to $\RR \times S^3,$ and
in the natural coordinates the metric has the form
\begin{equation}
ds^2 = dt^2 - \cosh^2(t) \, d\sigma^2.
\end{equation}
We recognize $dS$ as a special case of Robertson--Walker space--time with the
choice $S(t) = \cosh (t)$. Once again, we change time variables by defining
\begin{equation}  \label{eqtaut}
\tau = \arcsin(\tanh (t))
\end{equation}
(so that $d\tau/dt = 1/\cosh t$), which takes values $-\frac{\pi}{2} < \tau < \frac{\pi}{2}$.
Then the metric becomes
\begin{equation} \label{2}
ds^2 = \cosh^2 (\tau) \, \big(d\tau^2 - d\sigma^2\big).
\end{equation}
The infinite past is at $\tau = -\frac{\pi}{2}$ and the infinite future is at
$\tau = \frac{\pi}{2}$.
The isometry group of $dS$ is the de Sitter group $O(4,1)$,
{\it i.e.\ }it coincides with the Lorentz group on $\RR^5$.

     Comparing equations \eqref{1} and \eqref{2}, it is now clear
(and well--known
\cite{HE}) that each of the Robertson--Walker space--times specified
above can be conformally embedded into de Sitter space--time, {\it
  i.e.\ }there exists a global conformal diffeomorphism $\varphi$ 
(see Definition 9.16 in
\cite{BEE}) from $RW$ onto a subset of $dS$. How large the embedding
in $dS$ is depends on the range of the variable $\tau$ in each case examined,
which itself depends upon the function $S(t)$.
We will consider \RW as a submanifold of $dS$, equipped with both
the de Sitter metric $g$ and the Robertson--Walker metric $\grw$,
which are conformally equivalent:
\begin{equation} \label{eqgghat}
 \grw=\Omega^{-2}\,g\,,\mbox{ where }\;
\Omega(p) = \frac{\cosh(t(p))}{S(t(p))}.
\end{equation}

We shall now use these conformal embeddings to define
``wedges'' for these space--times in terms of ``wedges'' in de Sitter
space. A wedge in de Sitter space is  the causal
completion of the worldline of a freely falling observer. The
set of these wedges will be denoted by $\wdS.$
Similarly, we define wedges in \RW to be the intersections with \RW
of those de Sitter wedges whose 
edges\footnote{See below for a precise definition.}
are contained in $RW.$ They correspond to the causal
completions of the union of worldlines of freely falling observers
in $RW$.
\vspace*{-10mm}
\begin{figure}[ht] \label{F1}
\epsfxsize120mm
\epsfbox{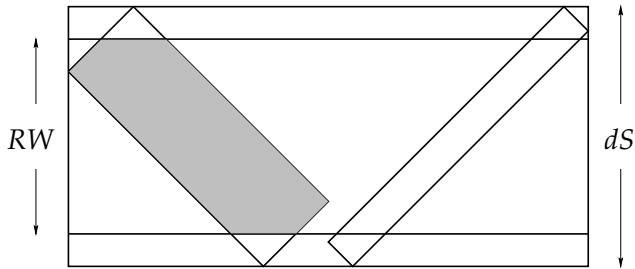}
\caption{Penrose diagram indicating a wedge in \dS (right) and in
  \RW (left).}
\end{figure}

The set of these
Robertson--Walker wedges will be denoted by $\wRW.$
It inherits useful properties from the family
$\wdS,$ which we collect in Lemma~\ref{LemWRW} and
which are most easily established with the following
alternative characterization of the de Sitter wedges.

Consider the
embedding~\eqref{eqdSMin} of \dS into Minkowski space and let
$\Lid$ denote the proper orthochronous Lorentz group in five dimensions.
Let then $\wMin$ be the family of regions obtained
by applying the elements of $\Lid$ to a single wedge--shaped region of
the form
\begin{equation}
\WRM \doteq \{ x \in \RR^5 \mid x_1 > \vert x_0 \vert \},
\end{equation}
\emph{i.e.\ }this family of regions is
$\wMin \doteq \{ \gammam \, \WRM \mid \gammam \in \Lid \}$.
Then $\wdS$ is just the collection of intersections
$\{ \wm \cap dS \mid \wm \in\wMin \}$.
There is clearly a one--to--one correspondence between $\wdS$ and $\wMin$.
For later convenience, we shall therefore denote by $\wm$ the
wedge in $\RR^5$ corresponding to a specified $W \in \wdS$.
With this characterization of $\wdS,$ one easily verifies that 
$\wRW$, the set of Robertson--Walker wedges, 
inherits the following  properties from $\wdS.$  These 
properties have been isolated in \cite{BDFS} as a distinctive feature
of wedge regions. 
\begin{lemma}\label{LemWRW}
 $\wRW$ is stable under the operation of taking causal
complements and under the action of the isometry group of
rotations $SO(4)$ on {\it RW}. Further, the elements of
$\wRW$ separate spacelike separated points in \RW and
$\wRW$ is a subbase for the topology in {\it RW}.
\end{lemma}
Note that with the preceding identification of spaces the action on
\RW of $SO(4)$ is just the restriction to \RW of its action on $dS$.

One obtains a more intrinsic characterization of $\wRW$ by noticing  
that wedges in de Sitter space--time can be characterized by their edges.
Let $\ERM$ be the edge of $\WRM$, {\it i.e.\ }the three--dimensional 
subspace $\{x \in \RR^5 :x_0=x_1=0 \}$.
Applying the elements of $\Lid$ to $\ERM$, one obtains all 
three--dimensional spacelike linear subspaces $\EM$ of \Min. The
intersections of these with \dS are exactly the
two--dimensional, spacelike, totally geodesic, complete, connected
submanifolds of \dS
(in other words, they are just 2-spheres \cite[p.105]{O}).
These submanifolds will be called
{\em  de Sitter edges} and are denoted by $E$. The causal complement
of $\ERM$ has two
connected components, one being $\WRM$ and the
other one being its causal complement $\WRM{}'\in\wMin.$ 
Hence also the causal complement of any de Sitter edge
has two connected components, each being a wedge,
{\it i.e.\ }a Lorentz transform of $\WRM$ intersected with 
$dS$. So we conclude that the
wedges in \dS may be characterized as the connected components of the
causal complements of de Sitter edges.

Based on this observation, we can
give an analogous, intrinsic characterization of wedges in a
Robert\-son--Wal\-ker space--time $RW$ after a pre\-pa\-ra\-tory lemma.
A submanifold $F$ of a semi--Riemannian
manifold $(M,g)$ is called  {\em totally umbilic} if there is a vector
field $Z_F$ normal to $F$ such that
\begin{equation}
  \nor_F \nabla_XY = g(X,Y)\,Z_F  ,
\end{equation}
for all vector fields $X,Y$ tangent to $F$. In this case, $Z_F$
is called the normal curvature vector field\footnote{Intuitively,
$F$ bends away from the normal curvature vector field if
$F$ is spacelike.}  of $F$. If, in particular,
$Z_F=0$, then $F$ is totally geodesic.

\begin{lemma} \label{Lem3.1}
Let $\frw$ be a submanifold of $dS.$ Then $\frw$ is a
totally geodesic submanifold of $(dS,g)$ contained in $RW,$
if and only if it is a totally umbilic submanifold of $(RW,\grw)$
with normal curvature vector field
\begin{equation} \label{3}
 \nor_{\frw} \big({\rm grad}(\ln \Omega)\big) .   
\end{equation} 
\end{lemma}

\nind Here the gradient $\grad f$ of a function $f$ denotes the vector field
which is metrically equivalent to the differential $df.$

\begin{proof}
Since the metrics $\grw$ and $g$ are conformally equivalent as
expressed in equation~\eqref{eqgghat}, the corresponding
connections  are related by
\begin{equation} \label{4}
\nrw_XY=\nabla_XY+\Omega\left(\,(X\Omega^{-1})Y+(Y\Omega^{-1})X-\grw(X,Y)
 \, \grad_{\grw}\, \Omega^{-1}\,\right) \,
\end{equation}
(see {\it e.g.\/} equation (2.29) in \cite{HE}). Now let $X,Y$ be vector fields
tangent to $\frw$, and denote the normal projections corresponding to $\frw$
with respect to $\grw$ and $g$ by $\nor_{\grw}$ and $\nor_g$,
respectively. Then the above equation implies
\begin{equation}
\nor_{\grw}\nrw_XY=\nor_{\grw}\nabla_XY +\grw(X,Y)\,
                            \nor_{\grw}\,\grad_{\grw} \ln\Omega.
\end{equation}
Taking into account the fact that $\nor_{\grw}\nabla_XY$ vanishes if and only
if $\nor_g\nabla_XY$ does, this proves the claim.
\end{proof}

Obviously,
$\frw$ is complete and spacelike w.r.t.\ the metric $\grw$ if and only if the
same holds for $g.$ Thus the set of de Sitter edges contained in
$\RWm$ coincides with the collection
of all two--dimensional, spacelike, totally umbilic, complete,
connected submanifolds of $(RW,\grw)$ with normal curvature vector fields
as in~\eqref{3}. These submanifolds  will be called {\em
Robertson--Walker edges}, denoted by $\erw$. The causal complement
of a Robertson--Walker edge $\erw$ coincides with the restriction
to $RW$ of its causal complement in $dS$ and  thus has two
connnected components, each being a wedge in $RW.$  Hence, the
set of wedges $\wRW$ in \RW can be intrinsically characterized as the
connected components of the causal complements of 
(the intrinsically defined) Robertson--Walker edges.

     We now exhibit a one--to--one map of the set $\wdS$ of de Sitter
wedges onto the set $\wRW$ of Robertson--Walker wedges,
which will be seen to have properties useful for our
``transplantation'' of nets performed in the next section.
Recall that an element of the class of Robertson--Walker space--times
considered here is embedded into \dS with a characteristic interval
$|\tau|< \alpha\frac{\pi}{2}\leq\pihalf$.
If $\alpha =1$, then the embedded $RW$ coincides with $dS$.
In this case a Robertson--Walker wedge is just a de Sitter wedge and
the families $\wdS$ and $\wRW$ are identified by the
embedding. To cover the general case $0 < \alpha \leq 1,$
we define a diffeomorphism $\Phi$ from
\dS onto $\RWm$ which bijectively maps the set of
de Sitter edges onto the set of Robertson--Walker edges.
It is given by
\begin{equation} \label{5}
 \Phi \left(\tau,\chi,\theta,\phi\right) \doteq  (f(\tau),\chi,\theta,\phi),
\end{equation}
where
\begin{equation} \label{6}
f(\tau) \doteq
 \arcsin\left(\sin(\alpha\pihalf)\,\sin(\tau)\right).
\end{equation}
The stated property of $\Phi$ as well as its uniqueness are established in
the Appendix. The map $\Phi$ gives rise to a one--to--one correspondence
\begin{equation}  \Xi: \wdS \rightarrow \wRW  \end{equation}
as follows. Let $W$ be a de Sitter wedge with edge $E.$
The causal complement  $\Phi(E)'$ of $\Phi(E)$ in $RW$ has two connected
components, exactly one of which has nontrivial intersection with $W$.

\begin{definition}
Let $W$ be a de Sitter wedge with edge $E.$
We define $ \Xi(W)$
to be the connected component of $\Phi(E)'$ in $RW$ which has
nontrivial intersection with $W$.
\end{definition}

\noindent
The map $\Xi$ thus maps the family of de Sitter wedges onto
the family of Robertson--Walker wedges. It has the following
specific properties which will be a key ingredient
in the transplantation of nets in the subsequent section.
\begin{prop} \label{Prop3.1}
The map $\Xi : \wdS \rightarrow \wRW$ is a
bijection.
It commutes with the operation of taking causal complements in the
respective spaces and
intertwines the action of the isometry group of rotations
SO(4) on $\wdS$ with its action on $\wRW$.  If $\alpha<1,$
then $\Xi$ is not induced by a bijective point transformation from
\dS onto $RW,$ {\em i.e.\ }there is no map $p:\dSm\rightarrow\RWm$ such
that for all $W\in\wdS$
\begin{equation*}
 \Xi(W) =\{p(x)\,|\,x\in W\}.
\end{equation*}
\end{prop}

The first part of this statement follows from the very construction of $\Xi$:
That $\Xi$ commutes with the operation of taking causal complements
is an immediate consequence of the
fact that any wedge has the same edge as its causal
complement. The intertwining properties of $\Xi$
are due to the fact that the action of $SO(4)$ commutes both with
the map $\Phi$ and with causal complementation. It is less obvious, however,
that $\Xi$ is not induced by a point transformation. This feature
originates from the fact that, though the scaling of edges is a
diffeomorphism, the subsequent causal complementation is not.
A formal proof of the latter statement as well as some further properties
of the map $\Xi$ are given below. The reader who wants to skip this quite
technical part may proceed at this point directly to the subsequent section.

As a first step in our analysis of $\Xi,$
we give a more explicit formula for its  action upon
$\wdS$. Denote by $\lambdam_2(t)$
the standard boost in $2$-direction of $\RR^5$, acting on the $0$- and
$2$- coordinates as
\begin{equation} \label{9}
\left( \begin{array}{cc}
 \cosh(t) &  \sinh(t)  \\
 \sinh(t) &  \cosh(t)
 \end{array} \right),
\end{equation}
and by $\lambda_2(\tau)$ the restriction of $\lambdam_2(t(\tau))$
to $dS$.
\begin{lemma} \label{Lem3.3}
Every wedge in \dS may be written as
\begin{equation} \label{10}
W = \rho\,\lambda_2(\tau)\, \wrds ,
\end{equation}
for some $\rho\in SO(4)$, $\tau\in(-\pihalf,\pihalf)$.
Similarly, every wedge in \RW may be written as the intersection of
such a wedge with $RW$, provided that $|\tau|<\alpha\pihalf.$
The bijection $\Xi: \wdS \rightarrow \wRW$ acts as
\begin{equation} \label{11}
  \Xi(\rho\,\lambda_2(\tau)\, \wrds) =
       \rho\,\lambda_2\big(f(\tau)\big)\, \wrds \, \cap \,\RWm.
\end{equation}
\end{lemma}

\begin{proof} There is a one--to--one correspondence between wedges $W$
in \dS and pairs of lightlike rays in \Min \ of the form
$\RR_+\,(\pm1,{\bf e}^\pm)$,
where $\mathbf{e}^\pm$ are unit vectors in Euclidean $\RR^4$
satisfying $\mathbf{e}^+\cdot\mathbf{e}^->-1.$
Namely, $\mathbf{e}^+$ and $\mathbf{e}^-$ are the unique unit vectors
such that 
$\Wtilde+(\pm1,\mathbf{e}^\pm) \subset\Wtilde.$ Conversely,
$\Wtilde$ is the set of all $(x^0,\mathbf{x})\in\Minm$\ satisfying
$-\mathbf{x}\cdot\mathbf{e}^-<x^0< \mathbf{x}\cdot\mathbf{e}^+.$
In particular, $\WRM$ corresponds to the rays $\RR_+\,(\pm1,1,0,0,0).$
The rays corresponding to $\lambdam_2(t)\, \WRM$ are calculated
to be
\begin{eqnarray}
& \lambdam_2(t)\,\RR_+\,(\pm1,1,0,0,0) =\RR_+\,(\pm1,\mathbf{f}^\pm(t)),
 & \nonumber \\
& \mathbf{f}^\pm(t) \doteq \big(\cosh(t)^{-1},\pm\tanh(t),0,0\big). &
\end{eqnarray}
Since $\mathbf{f}^+(t)\cdot\mathbf{f}^-(t)= 2\cosh(t)^{-2}-1$ exhausts
all values in $(-1,1]$ for $t\in\RR,$ one can fix $t$ such that
for $\mathbf{e}^\pm$ corresponding to a given wedge $W$, one has
$\mathbf{f}^+(t)\cdot\mathbf{f}^-(t)=\mathbf{e}^+\cdot\mathbf{e}^-.$
But then there exists a rotation
$\rhom\in SO(4)$ satisfying $\rhom\;\mathbf{f}^\pm(t)=\mathbf{e}^\pm.$
This shows that $W$ is of the form \eqref{10}.
Further, the intersection with \RW of a wedge of the form \eqref{10} is a
Robertson--Walker wedge if and only if its edge is contained in
$RW$. This is the case if and only if $|\tau|<\alpha\pihalf.$

It remains to prove relation \eqref{11} for the bijection
$\Xi: \wdS \rightarrow \wRW$.
 Denote by $\tilde{f}$
the map which results from $f,$ see equation \eqref{6},
under the coordinate transformation $\tau\rightarrow t:$ 
\begin{equation} \tilde{f}(t) \; \doteq \tau^{-1}\big(f(\tau(t))\big)\,
=\,\artanh\big(\sin(\alpha\pihalf)\tanh(t)\big).
\end{equation}
Recall that a point $x\in\Minm$ is in the edge $\ERM$ if and only if it
is of the form $x=(0,0,x_2,x_3,x_4)$. For such $x$, one calculates
(see definition \eqref{7} and equation ~\eqref{8} in the Appendix)
\begin{eqnarray}
& \tilde{\Phi} (\lambdam_2(t) x)
=
\big(\sin(\alpha\pihalf)\,\sinh(t)\,x_2,0,\cosh(t)\,x_2,x_3,x_4\,\big)
& \nonumber \\
&= \lambdam_2(\tilde{f}(t))\;
\big(\,0,0,\cosh(t) \, \cosh(\tilde{f}(t))^{-1} \,x_2,x_3,x_4\,\big). &
\end{eqnarray}
This shows by equation \eqref{8} that 
\begin{equation} \label{12}
\Phi (\lambda_2(\tau)\, \erds) =\lambda_2(f(\tau))\, \erds,
\end{equation}
where $\erds$ denotes the intersection of $\ERM$ with $dS$.
Now, the edge of the wedge $W$ in equation~\eqref{10} is
$ E = \rho\,\lambda_2(\tau)\, \erds.$ Since $\Phi$ commutes with the
rotations, equation~\eqref{12} implies
\begin{equation}
 \Phi(E) =  \rho\,\lambda_2(f(\tau))\, \erds.
\end{equation}
Obviously, $\rho\,\lambda_2(f(\tau))\, \wrds \, \cap \RWm$ is the
connected component of the causal complement (in \RW) of $\Phi(E)$ which has
nontrivial intersection with $W.$ This proves equation~\eqref{11}.
\end{proof}

We now discuss the behavior of intersections of wedges under the map
$\Xi.$  Recall that in Minkowski space 
a double cone is a nonempty intersection of a future lightcone
and a past lightcone. This definition makes sense in $dS$ and $RW$, as
well. We consider double cones whose future and past apices differ only in
the time coordinate: Let $x=(\tau,\chi,\theta,\phi)\in\dSm$ and let
$\eps$ be a positive number such that $|\tau\pm\eps|<\pihalf.$
To these data we associate a double cone
\begin{equation} \label{14}
\Os_{x,\eps} \doteq V_+(x_{-\eps})\cap V_-(x_{+\eps})\;,
\mbox{ where } x_{\pm\eps} \doteq (\tau\pm\eps,\chi,\theta,\phi).
\end{equation}
Here, $V_{\pm}(p)$ denotes the future respectively
past light cone with apex $p$.
Obviously $\Os_{x,\eps}$ is also a double cone in \RW 
if and only if $|\tau\pm\eps|<\alpha\pihalf,$ in which case it will be
denoted by $\dcrwsub{x,\eps}$.
\begin{prop}  \label{Prop4.2}
Let $x\in\RWm$ be arbitrary, and let $\eps$ be a positive number such that
$\dcrwsub{x,\eps}$ is contained in $RW.$ If $\eps> \pihalf(1-\alpha),$ then the
intersection of all wedges $W$ in \dS whose corresponding images 
$\Xi(W)$ in
\RW contain $\dcrwsub{x,\eps}$ is nonempty: it contains the \dS double cone
$\Os_{x,\epsc}$, where $\epsc = \pihalf-f^{-1}(\pihalf-\eps)$.
If, on the other hand, $\eps<\pihalf(1-\alpha),$ then the above
intersection of wedges is empty.
\end{prop}

The last statement implies in particular that, if $\alpha<1,$ there are wedges
$\wrwsub{i}\in\wRW$, $i=1,2,$ with nonempty intersection but with 
$\Xi^{-1}(\wrwsub{i})$ having empty intersection. Hence the map
$\Xi:\wdS\rightarrow \wRW$ cannot be induced by a bijective
point transformation, as claimed in Proposition~\ref{Prop3.1}.

For the proof of Proposition~\ref{Prop4.2} we need some further lemmas.
\begin{lemma}  \label{Lem4.1}
Let $\Os_{x,\eps}$ be a double cone as in
equation~\eqref{14} with
$x=(\tau_0,\pihalf,\pihalf,0).$ Let further $W$ be a wedge as in
equation~\eqref{10} with
$\rho=\rho'\,\rho_{14}(\omega'')\,\rho_{13}(\omega')\,\rho_{12}(\omega)$, where
$\rho_{1k}$ denotes a rotation in the $1$-$k$-plane in ambient
$\RR^5,$ $k=2,3,4,$ 
and $\rho'$ is a rotation which leaves the $1$-axis fixed.
Then $W$ contains $\Os_{x,\eps}$ if and only if the following two 
inequalities hold:
\begin{equation}
\cos(\omega'')\cos(\omega')\cos(\tau\pm\omega)\geq\cos(\pihalf-\eps\pm\tau_0).
\end{equation}
\end{lemma}
\begin{proof}
$W$ contains $\Os_{x,\eps}$ if and only if the apices of
$\lambda_2(-\tau)\rho^{-1}\,\Os_{x,\eps}$ are contained in the closure of
$\wrds,$ {\it cf.}\ \eqref{10}.  
The $0$- and $1$-components in ambient $\RR^5$ of these  two apices are
given by 
\begin{align}
\big(\lambda_2(-\tau)\rho^{-1}\,x_{\pm\eps}\big)_0 =&
\cosh(t)\sinh(t(\tau_0\pm\eps))  \nonumber\\
 & - b\,\sinh(t)\cosh(t(\tau_0\pm\eps))\sin(\omega)\,, \nonumber \\
\big(\lambda_2(-\tau)\rho^{-1}\,x_{\pm\eps}\big)_1 =& b\,\cosh(t(\tau_0\pm\eps))\cos(\omega),
\end{align}
respectively, where $t \doteq t(\tau)$ and
$b\doteq\cos(\omega')\cos(\omega'').$
Hence, taking into account the symmetry $t(-\tau)=-t(\tau)$, 
the two apices are contained in the closure of $\wrds$ if and only if
\begin{equation}
\tanh(t(\eps\pm\tau_0)) \leq
 b\left(\cosh(t)^{-1}\cos(\omega)\pm\tanh(t)\sin(\omega)\right).
\end{equation}
Since relation \eqref{eqtaut} between $t$ and $\tau$ implies
$\cosh(t(\tau))^{-1}=\cos({\tau}),$ these inequalities are equivalent to
\begin{equation}
\sin(\eps\pm\tau_0)\leq b\,\left(\cos(\tau)\cos(\omega)\pm
\sin(\tau)\sin(\omega)\right)
 =b\,\cos(\tau\mp\omega),
\end{equation}
which yields the assertions of the lemma.
\end{proof}

\begin{lemma}  \label{Lem4.2}
Let $W_1$ and $W_2$ be the wedges given by 
$W_i=$ $\rho_{12}(\omega_i)\,\lambda_2(\tau)\,\wrds,$ $i=1,2.$ If $\tau$ and
$\omega_- \doteq \half(\omega_1-\omega_2)$ satisfy
\begin{equation}  \label{31}
\cos(\omega_-)\,\cos(\tau+\omega_-)\leq 0,
\end{equation}
then $W_1\cap W_2=\emptyset$.
\end{lemma}

\begin{proof} As mentioned in the proof of Lemma~\ref{Lem3.3}, there are unique unit
vectors $\mathbf{e}^\pm\in\RR^4$ corresponding to a wedge $\wm$
such that $x=(x_0,\mathbf{x})\in \wm$ if and only if both
inequalities $\pm x_0<\mathbf{x}\cdot\mathbf{e}^\pm$ hold.
The unit vectors $\mathbf{e}_i^\pm$ corresponding to $\wm_i$
and hence to $W_i$ are determined by the equation
\begin{equation}
\RR_+\,(\pm1,\mathbf{e}_i^\pm) =
\rho_{12}(\omega_i)\,\lambdam_2(t(\tau))\,\RR_+\,(\pm1,1,0,0,0)
\end{equation}
to be
\begin{equation}
\mathbf{e}_i^\pm =
(\,\cos(\tau\mp\omega_i)\,,\,\pm\sin(\tau\mp\omega_i)\,,\,0\,,\,0).
\end{equation}
Let now $x\in W_1\cap W_2 \subset \wm_1 \cap \wm_2$. Then
$x$ must satisfy
\begin{equation}
0< \,\mathbf{x}\cdot
 (\mathbf{e}_1^++\mathbf{e}_1^-+\mathbf{e}_2^++\mathbf{e}_2^-)=
4\cos(\tau)\cos(\omega_-)\big(x_1\cos(\omega_+)-x_2\sin(\omega_+)\big)
\end{equation}
and
\begin{equation}
0< \,\mathbf{x}\cdot (\mathbf{e}_1^-+\mathbf{e}_2^+)=
 2\cos(\tau+\omega_-)\big(x_1\cos(\omega_+)-x_2\sin(\omega_+) \big),
\end{equation}
where $\omega_+ \doteq \half(\omega_1+\omega_2).$
It follows that $\cos(\omega_-)$ and $\cos(\tau+\omega_-)$ are non--zero
and have the same sign, since $\cos(\tau)>0$ by assumption.
This contradicts~\eqref{31}. 
\end{proof}

     We can now prove Proposition~\ref{Prop4.2}.

\medskip

{\noindent \bf Proof of Proposition \ref{Prop4.2}.} By $SO(4)$ covariance,
it suffices to consider the special case 
$x=(\tau_0,\pihalf,\pihalf,0).$ First let 
$\eps>\pihalf(1-\alpha)$ and let
$\dcrwsub{x,\eps}\subset\wrw$, where 
$\wrw=\rho\lambda_2(\tau) \wrds \cap\RWm$, 
with $|\tau|<\alpha\pihalf.$  By Lemma~\ref{Lem4.1},
this is equivalent to the two conditions for $+$ and $-,$
respectively, 
\begin{equation} \label{15}
 -\arccos\big( b^{-1} \cos(\eps'\pm\tau_0)\big)\leq\tau\pm\omega\leq
 \arccos\big( b^{-1} \cos(\eps'\pm\tau_0)\big),
\end{equation}
where $\eps' \doteq \pihalf-\eps$ and $b \doteq \cos(\omega')\cos(\omega'').$
It must be shown that $W = \Xi^{-1}(\wrw)$ contains $\Os_{x,\epsc}$ with
$\epsc=\pihalf-f^{-1}(\pihalf-\eps).$ According to Lemma~\ref{Lem4.1}, this is the case
if and only if the two above conditions hold with $\tau$ replaced by
$f^{-1}(\tau)$ and $\eps'$ replaced by
\begin{equation}
\epsc' \doteq \pihalf-\epsc= f^{-1}(\eps').
\end{equation}
First consider the function $h(x) \doteq \arccos(b^{-1}\cos(x))$ for
$x\in(0,\pi)$.  Then $h(x)-x$ is an increasing function because
$h'(x)=\sin(x)\big(b^2-\cos^2(x)\big)^{-1/2}\geq1$. Since
$\eps'\leq f^{-1}(\eps')$ and $\eps'\leq\pihalf$, this entails the relations
\begin{equation} \label{16}
 h(\eps'\pm\tau_0) -\eps' \leq \;
 h(f^{-1}(\eps')\pm\tau_0) -f^{-1}(\eps')
\end{equation}
and
\begin{equation}
 h(\eps'+\tau_0) -\eps' \leq \;h(\pi-\eps'+\tau_0)- \pi+\eps'.
\end{equation}
But $h(x)=\pi-h(\pi-x)$, so the right--hand
side of the latter inequality coincides with $-h(\eps'-\tau_0)+\eps'$
and consequently $h(\eps'+\tau_0)+h(\eps'-\tau_0) \leq 2\eps'$.
Hence, by adding the inequalities \eqref{15} corresponding to ``$+$'' to those
corresponding to ``$-$'', one obtains
\begin{equation}  \label{18}
 -\eps'\;\leq\;\tau\;\leq\; \eps'.
\end{equation}
Recall that $f^{-1}(x)=\arcsin\big(\sin(\alpha\pihalf)^{-1}\sin(x)\big)$ with
domain $(-\alpha\pihalf,\alpha\pihalf)$. This function is odd and has
derivative
$\cos(x)\big(\sin^2(\alpha\pihalf)-\sin^2(x)\big)^{-1/2}\geq1.$
Hence $f^{-1}(x)-x$ is an increasing function, and \eqref{18} implies
\begin{equation} \label{19}
-f^{-1}(\eps')+\eps'\;\leq\;f^{-1}(\tau)-\tau\;\leq\;f^{-1}(\eps')-\eps'.
\end{equation}
Combining \eqref{19}, \eqref{16} and the assumption \eqref{15} yields
\begin{equation}
 -\arccos\big( \frac{1}{a}\cos(f^{-1}(\eps')\pm\tau_0)\big) \;\leq\;
f^{-1}(\tau)\pm\omega\;\leq\;
\arccos\big(\frac{1}{a}\cos(f^{-1}(\eps')\pm\tau_0)\big),
\end{equation}
and this shows that $W$ contains $\Os_{x,\epsc}.$

     Let now $\eps<\pihalf(1-\alpha),$ {\it i.e.\ } 
$\eps' =\pihalf-\eps>\alpha\pihalf.$ The goal is to exhibit wedges
$W_1,W_2$ with empty intersection but satisfying
$\wrwsub{1}\cap\wrwsub{2} \supset\dcrwsub{x,\eps}.$ To this end, let for
$i=1,2$
\begin{equation}
 \wrwsub{i} (\delta) \doteq
    \rho_{12}(\omega_{i,\delta})\,\lambda_2(\tau_\delta)\; \wrds
\cap \RWm,
\end{equation}
where $\delta\in (\,0\,,\,\alpha\pihalf-|\tau_0|\,)$ and
\begin{equation}
\tau_\delta=\alpha\pihalf-\delta\;,\quad
\omega_{1,\delta}=\tau_0+\eps'-\tau_\delta\;,\;\mbox{ and } \;
\omega_{2,\delta}=\tau_0-\eps'+\tau_\delta.
\end{equation}
For $\delta$ in the specified range, Lemma~\ref{Lem4.1} asserts that
$\wrwsub{1} (\delta)\cap\wrwsub{2} (\delta) \supset\dcrwsub{x,\eps}$. On the
other hand, the $W_i(\delta)$ are given by
$\rho_{12}(\omega_{i,\delta})\,\lambda_2(f^{-1}(\tau_\delta))\;\wrds,$
cf.\ Lemma~\ref{Lem3.3}. Now for all admissible $\delta$, one has
\begin{equation}
 0\,<\, \half(\omega_{1,\delta}-\omega_{2,\delta})
   =\eps'-\alpha\pihalf+\delta \,<\,\eps'\,<\,\pihalf.
\end{equation}
Further, the expression
$f^{-1}(\tau_\delta)+\half(\omega_{1,\delta}-\omega_{2,\delta})$ 
is continuous in $\delta$ and approaches the value
$\pihalf+\eps'-\alpha\pihalf>\pihalf$ if $\delta$ tends to zero. Hence for some
$\delta_0>0$ this expression is greater $\pihalf$.
But then Lemma~\ref{Lem4.2} entails 
$W_1(\delta_0)\cap W_2(\delta_0)=\emptyset.$ \hfill $\square$

The action of the proper de Sitter group $SO(4,1)$ on \dS induces an action
on $\wdS.$ Via $\Xi,$ one has then an action of $SO(4,1)$ on the set
of Robertson--Walker wedges $\wRW$ given by
\begin{equation}  \label{eqgW} 
   \gammarw \, \wrw \doteq (\Xi \circ \gamma \circ\Xi^{-1})(\wrw).
\end{equation}
We finally discuss the question of whether  this action is induced by point
transformations on $RW,$ {\it i.e.\ }if for $\gamma\in SO(4,1)$ there
exists a map $p_\gamma:\RWm\rightarrow \RWm$
such that
\begin{equation} \label{eqPTg}
 \gammarw \, \wrw = \{p_\gamma (x)\,|\,x\in \wrw \}.
\end{equation}
\begin{prop} \label{PropPTg}
For the subgroup $SO(4)$ of rotations the action~\eqref{eqgW} on
$\wRW$ is induced by its  action on $RW$ as (isometric) point
transformations. However if $\alpha<1,$ then there are
elements in $SO(4,1)$ which are not induced by a
point transformation in the sense of equation~\eqref{eqPTg}.
\end{prop}
\begin{proof}
The first statement has been established already in
Proposition~\ref{Prop3.1}, while the latter one is a
consequence of the subsequent Lemma~\ref{Lem4.3}. 
\end{proof}

Note that $|f(\tau)| < |\tau|$ if $\alpha < 1$. Hence, in this case
it is possible to find $\omega\in (0,\pihalf)$ and $\tau>0$ satisfying
\begin{equation} \label{20}
f(\tau)+\omega<\pihalf\;,\;\tau+\omega>\pihalf.
\end{equation}
For such $\omega$ and $\tau$, $\cos(\omega)$ is smaller than
$\cos(\pihalf-\tau)=\sin(\tau)$, so that $\tau_0$ is well defined by
\begin{equation} \label{21}
 \sin(\tau_0) = -\cos(\omega)\sin(\tau)^{-1} .
\end{equation}

\begin{lemma}  \label{Lem4.3}
Let $\alpha < 1$ and let
$W_\pm \doteq \rho_{12}(\pm\omega)\,\lambda_2(\tau)\, \wrds$, where
the parameters $\omega\in (0,\pihalf),$ $\tau>0$ satisfy \eqref{20}.
Consider the de Sitter isometry $\gamma_0 \doteq \lambda_2(\tau_0),$
with $\tau_0$  as above.
Then $\wrwsub{+}\cap\wrwsub{-}\neq\emptyset,$ but
$\gammarw_{\!0} \, \wrwsub{+}\cap \gammarw_{\!0} \, \wrwsub{-}=\emptyset.$
\end{lemma}

\begin{proof}
Let $\omega$ and $\tau$ satisfy \eqref{20}. Then Lemma~\ref{Lem4.2} entails
${W_+}\cap{W_-}=\emptyset$. Moreover, the intersection of the wedges
$\wrwsub{\pm}= \rho_{12}(\pm\omega)\,\lambda_2(f(\tau))\,\wrds \cap \RWm$ is
non--empty, because by Lemma~\ref{Lem4.1} it contains a double
cone $\dcrwsub{x,\eps},$ where $x=(0,\pihalf,\pihalf,0)$ and $\eps>0.$
For the proof of the second part of the statement we proceed to
ambient Minkowski space and denote by $\gammam$ the Lorentz
transformation corresponding to a given de Sitter transformation
$\gamma.$ Since $\lambdam_2(\cdot)$ and $\rhom_{12}(\cdot)$ act only on the
$x_0,x_1$ and $x_2$-coordinates of $\WRM,$ it follows from the
argument in Lemma~\ref{Lem3.3} that there are $\omega',\tau'$ such that 
\begin{equation} \label{50}
 \gammam_0 \, \wm_+=\rhom_{12}(\omega')\,\lambdam_2(\tau')\, \WRM.
\end{equation}
Further, a Lorentz transformation acting only on $x_0,x_1$ and $x_2$
leaves $\WRM$ invariant if and only if it leaves the unit vector in the
$x_2$-direction $e_2$
invariant. Hence, $\omega'$ and $\tau'$ satisfy the above equation
if and only if they satisfy the condition
\begin{equation} \label{22}
\lambdam_2(\tau_0)\,\rhom_{12}(\omega)\,\lambdam_2(\tau)\,e_2=
\rhom_{12}(\omega')\,\lambdam_2(\tau')\,e_2,
\end{equation}
which implies
\begin{equation}
\cot(\omega')= \big(\sin(\tau_0) \sin(\tau)+\cos(\omega)\big) 
\big(\sin(\omega) \cos(\tau_0)\big)^{-1}.
\end{equation}
The reflection $\jm_2$ about the edge   
$\{ x \in \RR^5: x_0 = x_2 = 0 \}$
commutes with $\lambdam_2(\cdot),$ satisfies
$\jm_2\,\rhom_{12}(\omega)=\rhom_{12}(-\omega)\,\jm_2$ 
and maps $\WRM$ onto itself. 
Hence, applying $\jm_2$ to equation \eqref{50}, it follows that
$\gammam_0 \, \wm_-=\rhom_{12}(-\omega')\,\lambdam_2(\tau')\, \WRM$. 
Combining the preceding  facts one gets
$\gammarw_{\!0} \, \wrwsub{\pm}=\rho_{12}(\pm\omega')\,\lambda_2(f(\tau'))\, \wrds
\cap \RWm$.
But $\cos (\omega')=0$ for $\tau_0$ given by equation~\eqref{21}, hence
$\gammarw_{\!0} \, \wrwsub{+}\cap\gammarw_{\!0} \, \wrwsub{-}=\emptyset$ by 
Lemma~\ref{Lem4.2}.
\end{proof}

This completes our discussion of the properties of the map $\Xi$.

\section{Transplantation of Nets and States}
We now turn to the construction of theories on Robertson--Walker
space--times from theories on de Sitter space.
Thus we commence with a net of algebras on $dS$ and a
corresponding state and define an
associated net and state on $RW$. We shall
see that this {\em transplantation} of net and state preserves the
properties of causality and $SO(4)$ covariance. Moreover, the transplanted
theory satisfies the CGMA and a \msc \, if the original theory does. We
shall also investigate under which  conditions
the action of the modular symmetry group upon
the transplanted net is induced by point transformations on the
corresponding space--time.

For the  convenience of the reader we recall the general formulation of
the CGMA and the \msc.
Let $M$ be a space--time and $\wedges$ be a suitable collection of
open subsets of $M.$
Let further $\wnet$ be a net of $C^*$--algebras indexed by $\wedges$,
each of
which is a subalgebra of a $C^*$--algebra $\As$. A state on $\As$ will be
denoted by $\omega$ and the corresponding GNS representation of $\As$ will be
signified by $(\Hs,\pi,\Omega);$ $\omega$ will also be said to be a state on
the net $\wnet$. For each $W \in \wedges$ the von Neumann algebra $\pi(\As(W))''$
will be denoted by $\Rs(W)$.           The modular involution
associated by Tomita--Takesaki theory~\cite{BratRob}
to the pair $(\Rs(W),\Omega)$ --- whenever $\Omega$ is
cyclic and separating for $\Rs(W)$ --- will be represented by $J_W$, while
the unitary modular group associated to the same pair will be written as
$\{\Delta_W^{it}\}_{t \in \RR}$. We emphasize that
these modular objects are uniquely determined by the algebraic data.

\begin{definition}[Condition of Geometric Modular Action]
Let $\wedges$ be a suitable collection of open subsets of the space--time
$M$, let $\wnet$ be a net of $C^*$--algebras indexed by $\wedges$, and let
$\omega$ be a state on $\wnet$. The CGMA is fulfilled if the corresponding net
$\wrnet$ satisfies \\[1mm]
(i) $\;W \mapsto \Rs(W)$ is an  order preserving bijection \\[1mm]
(ii) $\,\Omega$ is cyclic and separating for $\Rs(W),$ 
for each $W \in \wedges$ \\[1mm]
(iii) the adjoint action of $J_{W_0}$
leaves the set $\wrnet$ invariant,  for each $W_0 \in \wedges$.
\end{definition}

\vspace*{1mm}
See \cite{BDFS} for a motivation for and consequences of this condition 
and also for a discussion of the meaning of
``suitable collection''. For the class of Robertson--Walker space--times
considered here and for de Sitter space--time, the respective
wedge regions $\wdS, \wRW$ are suitable in this sense, {\it cf.}\
Lemma~\ref{LemWRW}. 

Given a net and state satisfying the CGMA, one is faced with the
physically important question of the stability of  the state.
For Minkowski space theories this stability is
usually characterized by the relativistic spectrum condition, {\it
  i.e.\ }the
joint spectrum of the generators of the representation of the translation
subgroup is contained in the closed forward light cone \cite{SW,Haag}.
However, such a subgroup is not to be found in the isometry group of most
space--times of physical interest.
  For this reason, work has been invested in finding a
replacement for the relativistic spectrum condition as a stability
condition valid for general space--times. We mention, in particular,
the interesting microlocal spectrum condition
\cite{Rad1,Rad2,BFK,V}. In \cite{BDFS}, an alternative has been
suggested which relies on the modular objects.

\begin{definition}[Modular Stability Condition] The modular unitaries are
contained in the group $\Js$ generated by the modular involutions
$J_W,$ $W\in\wedges,$ {\em i.e.}
$\Delta_W^{it} \in \Js$, for all $t \in \RR$ and $W \in \wedges$.
\end{definition}

     This condition is formulated without any reference to an
underlying space--time; hence, it may be applied in principle to models
on any space--time. Though it is certainly not clear {\it prima facie}
that the \msc \, has anything to do with physical stability, it
was shown in \cite{BDFS} that in Minkowski space theories, the CGMA
and the \msc \, {\it entail} the relativistic spectrum condition.

In the present context it is of interest that
there  exist many models on de Sitter space
satisfying the CGMA and the \msc.
We recall that these properties follow, within de Sitter covariant
theories, from a characterization of vacuum states via a KMS
condition~\cite{BB}, {\it cf.\ }also \cite{BEM}.
 The latter condition is satisfied, for example, by the (generalized) free
field models constructed on de Sitter space in~\cite{BM} and \cite{NPT}.
Moreover, as proposed in~\cite{Fred}, a suitable theory on ambient Minkowski
space can be restricted to \dS in a specific way such that the
resulting theory has, by the Bisognano--Wichmann
theorem~\cite{BW}, the above properties.

We now start with such a de Sitter theory: Let $\wnetds$  and $\omega$
be the corresponding net and state, {\it i.e.\ }we make the choice
$\wedges =\wdS,$ the set of de Sitter wedges. The group $\Js$ generated by the
modular involutions  then provides a
continuous (anti)unitary representation of the proper de Sitter group
$SO(4,1),$ under which the net $\wrnetds$ is covariant~\cite{BB}. 
Moreover, the net satisfies wedge duality and thus is
local. We proceed from the given net on $dS$ to a corresponding net
$\wnetrw$ on $RW$ by {\it transplantation}, putting
\begin{equation}
\Arw(\wrw) \doteq \As(W),\quad
\wrw = \Xi(W),
\end{equation} 
where $\Xi:\wdS\rightarrow \wRW$ is the bijection defined in the
 preceding section. In addition, we proceed from $\omega$
to a corresponding state $\orw$ on $\wnetrw$ by
\begin{equation}
\orw(A) \doteq \omega(A) \quad \textnormal{for all} \quad
    A \in \Arw(\wrw), \ \wrw \in \wRW. 
\end{equation}
We thus obtain a net $\wnetrw$ and state
$\orw$
on $RW$, which coincide in the indicated manner with the net $\wnetds$
and $\omega$, but which now are re-interpreted in terms of Robertson--Walker
space--time. The physical significance of the operators and state
thereby changes.

The modular symmetry group $\Js$ induces an action of the de Sitter
group $SO(4,1)$ on the Robertson-Walker net, as one easily verifies.
More specifically, for any $\gamma\in SO(4,1)$ there is
a unique $J\in\Js$ such that
\begin{equation} \label{eqJRJ}
J\Rrw(\wrw)J^{-1} = \Rrw ({\gammarw}\wrw) \quad
\text{for all} \quad \wrw \in \wRW,
\end{equation}
where $\gammarw \wrw = (\Xi \circ \gamma \circ \Xi^{-1})(\wrw)$
is the action defined in equation~\eqref{eqgW}.
It was shown in Proposition~\ref{PropPTg} that
this action on $\wRW$ is induced by an isometry of \RW if
$\gamma\in SO(4),$ but, for arbitrary $\gamma\in SO(4,1),$ need not even
arise from a point transformation if $\alpha<1$.
\begin{theorem} \label{Thm4.1}
With the above definitions, the net $\wnetrw$ and state
$\orw$ satisfy the CGMA and
the \msc. The corresponding net of von Neumann algebras
$\wrnetrw$ satisfies wedge duality
and transforms, in the sense of equation \eqref{eqJRJ},
covariantly under $SO(4,1)$  under the adjoint action of $\Js$.
In particular, the group $SO(4)$ of isometries of
\RW is implemented by a subgroup of $\Js$. 
On the other hand, for temporal scaling factor
$\alpha < 1$, there exist also $J \in \Js$ which do not implement
point transformations of $RW$.
\end{theorem}
\begin{proof} The claimed properties follow
from the corresponding properties of the underlying de Sitter net,
taking into account the specific features of $\Xi$ established in
Section~2. 
\end{proof}

It is of interest at this point to consider the special case of
Robertson--Walker spaces for which the function $S$ is such that the
temporal scaling parameter $\alpha = 1$. In this case, the
conformal diffeomorphism $\varphi:\RWm\rightarrow \dSm$ is onto, thus
establishing a conformal equivalence of \RW and $dS.$ Hence the
conformal groups of the two space--times are isomorphic. But the
conformal group of \dS coincides with its isometry group, the de
Sitter group $O(4,1).$ This follows from the characterization
in~\cite{Les3} of the de Sitter group as the bijections of \dS which
preserve separation zero.
Hence, the conformal group of \RW is isomorphic to the de Sitter
group, which acts on \RW via the conformal embedding, {\it i.e.}
$\gamma\in 0(4,1)$ acts on \RW as $\varphi^{-1}\circ \gamma \circ \varphi.$
Taking into account that the map $\Xi$ is induced by $\varphi$ in
these cases, we arrive at the following statement.
\begin{cor} \label{ThmAlpha1}
With the above definitions and for Robertson--Walker
space--times $RW$ with temporal scaling factor $\alpha = 1,$
the net $\wrnetrw$  satisfies, in addition to the results of
Theorem~\ref{Thm4.1}, conformal covariance. More precisely, $\Js$
provides a representation of the group of conformal orientation
preserving transformations $SO(4,1)$
of $RW,$ under
whose adjoint action the net $\wrnetrw$ is covariant.
\end{cor}

We now turn to the analysis of algebras associated with precompact
subsets of $RW$ such as double cones.
For double cones $\dcrw \subset RW$, we define
\begin{equation} \label{56}
\Rrw (\dcrw) \doteq \; \bigcap_{\wrw \supset
\dcrw}\, \Rrw (\wrw) .
\end{equation}
These are the maximal algebras one can associate to
double cones such that the resulting net including these double cone algebras
fulfills the condition of isotony. It is an
immediate consequence of wedge duality
that the so--defined net is local.

    Though  the $\Rrw (\dcrw)$ defined in \eqref{56} are the largest algebras
one can associate to double cones $\dcrw$ in a meaningful way,
it may be that some or all of them are
trivial. This is a matter of some significance. If all double cone
algebras are trivial, the net is physically irrelevant, since then no
observable can be localized in any bounded spacetime region even though all
{\it real} observations are necessarily so localized. If,
however, algebras associated with sufficiently large double cones are
non--trivial but those associated with sufficiently small double cones
are trivial, then the net describes a system for which there is a
length scale below which no observables can be localized. In fact,
for the net $\{ \Rrw (\dcrw) \}_{\dcrw\subset RW}$
this question is settled by the following corollary, which is an
immediate consequence of Proposition~\ref{Prop4.2} if the underlying 
net on \dS complies with the condition of weak additivity \cite{BB} and 
if intersections of algebras corresponding to disjoint wedges are
trivial \cite{TW}.
\begin{cor} \label{Cor4.1'}
Let $x\in\RWm$ be arbitrary, and let $\eps$ be a positive number
such that\,\footnote{Recall the definition \eqref{14} of the
  double cones $\dcrwsub{x,\eps}.$ } $\dcrwsub{x,\eps}$
is contained in $RW.$
If $\eps> \pihalf(1-\alpha)$, the GNS-vector $\Orw$ representing $\orw$
is cyclic for $\Rrw(\dcrwsub{x,\eps})$, whereas if
$\eps<\pihalf(1-\alpha)$, then $\Rrw(\dcrwsub{x,\eps}) = \CC\cdot 1$.
\end{cor}
     
Note that $\dcrwsub{x,\eps}$ can be contained in \RW only if
$\eps<\alpha\pihalf.$ In this case, $\alpha<\half$ implies
$\eps<\pihalf(1-\alpha).$ Hence, only if $\alpha\geq\half$ are 
there sufficiently large double cones
$\dcrwsub{x,\eps}$ such that the associated algebras are non--trivial.

     Summarizing, we have shown that for any value of the temporal scaling
factor $0 < \alpha \leq 1$, the CGMA and the \msc \, are satisfied
by the net $\wrnetrw$ and the state $\orw$.
When the scaling factor $\alpha$
equals $1$, $\Js$ provides a representation of the conformal group $SO(4,1)$
of $RW$ under  whose adjoint action the net is covariant.
If $\alpha < 1$, $\Js$ still induces a geometric action of the
group $SO(4,1)$ on the net, but this action can in general not 
be interpreted in terms of point transformations on $RW$.

It seems plausible  that similar results can be obtained for a
broader class of Robertson--Walker space--times. In fact, all
Robertson--Walker space--times can be conformally embedded into the
Einstein static universe \cite{HE}, on which one can find well-behaved
nets and states~\cite{D}. It should be possible to promote these
 desirable properties to the Robertson--Walker spaces by
transplantation. More immediately, a class of
Robertson--Walker spaces with negative curvature can be conformally
embedded into Minkowski space \cite{HE}.
In addition, Robertson--Walker spaces with 0 sectional curvature and
corresponding to the $\alpha = 1$ case are globally conformally
equivalent to Minkowski space. This provides the opportunity to verify
results analogous to those worked out  above.
\section{Concluding Remarks}
Making use of the novel method of transplantation of nets, we have
been able to make two things clear. First
of all, the CGMA is by no means limited in its scope of application to
space--times with maximal isometry groups, like
the special cases of Minkowski space and de Sitter space worked out in
\cite{BDFS,F2,BFS2}. Second and closely related, the group
$\Js$ arising from the CGMA can be much larger than the symmetry group
of the underlying space--time.  This supports our view that the CGMA,
coupled with the \msc, could be a useful selection criterion even for
space--times with trivial symmetry groups, which, after all, is the
generic case.

     A further point of interest is that the \msc \, holds in the
examples presented in this paper.  We recall that the \msc \, can
be expressed for theories on arbitrary space--times. Moreover, for
Minkowski space the CGMA and the \msc \, together yield the usual
condition of physical stability --- {\it i.e.}\ the
relativistic spectrum condition. Recent results concerning theories in
de Sitter space \cite{BEM,BB} are consistent with our proposal
of the \msc \, as a criterion for stable  vacuum--like systems.
Indeed, there are indications
\cite{Fred2} of a possible connection between the \msc \, and the
microlocal spectrum condition for non--Minkowskian space--times.

     Since the existence of nets and states on certain
Robert\-son--Walker space--times which satisfy the CGMA
and the \msc \, has now been established, it seems rewarding to
determine in a next step whether the symmetry group of these
space--times along with a corresponding (anti)unitary
representation can be recovered from a net and state
satisfying these conditions.
In this context, it would also be interesting to determine if the
net \cite{D} and adiabatic vacuum states~\cite{LR} (or the adiabatic KMS
states constructed in \cite{Tr}) associated with the free scalar
Bose field satisfy the CGMA and the \msc \, for
our choice of wedges $\wRW$. We hope to return elswhere to these
intriguing aspects of the general program outlined in \cite{BDFS}.

\appendix
\section{Appendix}  
\begin{lemma} \label{Lem3.2}
Let $\Phi$ be the diffeomorphism from \dS onto $\RWm$
given by equations \eqref{5} and \eqref{6}. Then $\Phi$ induces a
bijection from the  set of de Sitter edges onto the set of
Robertson--Walker edges. 
\end{lemma}
\begin{proof} The first step is to show that $\Phi$ arises from the linear map
$\tilde{\Phi}$  in \Min \ defined by
\begin{equation} \label{7}
 \tilde{\Phi}(x_0,\mathbf{x})\doteq\big(\sin(\alpha\pihalf)\,x_0,\mathbf{x}\big).
\end{equation}
Since $0<\sin(\alpha\pihalf)\leq 1,$ $\tilde\Phi$ leaves the set of
spacelike vectors invariant. Denoting $|y| \doteq (-y^2)^\half$ for
spacelike $y,$  the claim is that
\begin{equation}  \label{8}
\Phi(x)=
{\tilde{\Phi}(x)}\,{|\tilde{\Phi}(x)|^{-1}}.
\end{equation}
To see this, let $\Phi_1$ be the map defined by the  right hand side of the
above equation. Recall~\cite{HE} that the $t$-coordinate of a point
$x=(x_0,\mathbf{x})\in\dSm$ is given by
\begin{equation}
\tanh(t(x)) = {x_0}\,{\|\mathbf{x}\|^{-1}},
\end{equation}
where $\|\mathbf{x}\|$ denotes the Euclidean norm of $\mathbf{x},$
and this expression coincides with $\sin(\tau(x))$ by
equation~\eqref{eqtaut}. Furthermore,
the coordinates $(\chi,\theta,\phi)(x)$ are just the natural
$S^3$ coordinates of $\|\mathbf{x}\|^{-1}\mathbf{x}$. Thus one easily
verifies that the $S^3$-coordinates are left invariant by
$\Phi_1,$ while the $\tau$-coordinate transforms according to
\begin{equation}
\sin(\tau(\Phi_1(x)) = \sin(\alpha\frac{\pi}{2}) \sin(\tau(x)).
\end{equation}
Hence, $\Phi_1$  coincides with $\Phi,$ proving equation
\eqref{8}. But this equation implies that for any spacelike
linear subspace $\LM\subset\Minm$
\begin{equation}
\Phi(\LM \cap \dSm) = \tilde\Phi(\LM) \cap \dSm.
\end{equation}
It follows that $\Phi$ leaves invariant the set of intersections of \dS with
three-dimen\-sion\-al spacelike linear subspaces of \Min, {\it i.e.}\ the
set of de Sitter edges.
Now let $\LM$ be a spacelike linear subspace of \Min \ whose
intersection with \dS is contained in $\RWm$. Then the
$\tau$-coordinates of the intersection are contained in the  interval
$(-\alpha\pihalf,\alpha\pihalf).$ Hence every $x\in \LM$ satisfies
$|x_0| \, \|{\mathbf x}\|^{-1}<\sin(\alpha\pihalf)\leq1,$ and therefore
$\tilde{\Phi}^{-1}(x)= (\sin(\alpha\pihalf)^{-1}x_0,{\mathbf x})$ is
spacelike. This shows that the preimage under $\Phi$ of every edge
contained in $\RWm$ is a de Sitter edge.
\end{proof}

We now turn to the question of uniqueness. Let $\Phi'$ be a second
diffeomorphism from \dS onto $\RWm$ which bijectively maps the set of
de Sitter edges onto the set of Robertson--Walker edges.
Then one concludes that $\Phi_0\doteq \Phi^{-1}\circ \Phi'$ maps the
set of spacelike geodesics in $dS$ onto itself. 
Since any pair of lightlike separated 
points in \dS can be approximated by a pair of spacelike separated
ones, one thus verifies that $\Phi_0$ preserves separation zero in
both directions. But then $\Phi_0$ is an isometry of $dS,$ as shown
in~\cite{Les3}. Since there is no isometry $\Phi_0\in O(4,1)$ which
acts only on the time variable other
than the identity and time reversal, we have the following result.
\begin{lemma} The map $\Phi$ is, up to isometries of $dS$,
the only diffeomorphism from \dS onto $\RWm$ which maps de Sitter
edges to Robertson--Walker edges. Furthermore, up to time reversal, it is the
only such map which acts only on the time variable in the chosen
coordinate system.
\end{lemma}

\vspace*{2mm}
\noindent {\Large \bf Acknowledgements} \\[2mm]
We gratefully acknowledge financial support
from the Deutsche Forschungsgemeinschaft (DFG).

\end{document}